\documentclass[11pt]{article}
\usepackage{homework} %formatting the homework
\usepackage{graphicx}

\title{Product/Brand extraction from WikiPedia}
\date{}
\begin{document}

\maketitle
\begin{abstract}
In this paper we describe the task of extracting product and brand pages from wikipedia. We present an experimental environment and setup built on top of a dataset of wikipedia pages we collected. We introduce a method for recognition of product pages modelled as a boolean probabilistic classification task. We show that this approach can lead to promising results and we discuss alternative approaches we considered.
\end{abstract}

%\tableofcontents

%\newpage

\section{Introduction}
The aim of this work is to extract product and brand pages from the wikipedia corpus. Heuristically, we call a product anything that can be bought and for which a price can be determined. The definition of brand follows either as a family (line) of products or as the name of a manufacturer. At this stage we did not consider services (ie: twitter, google) as products. Several approaches have been carried on to perform this task and will be discussed in this paper. The solution we propose is to model the extraction process in the fashion of a classification problem. Given the wikipedia corpus we created a training set consisting of products and brands and we trained a Naive Bayes Classifier(NBC) to recognize unseen instances of wiki pages.

In the first part of the paper we will discuss related work that inspired our approach. Following we introduce a data set of wikipedia pages that we collected
and present an experiment methodology. In \textit{Classification} section we describe a probabilistic classification method we used to categorize pages and discuss results obtained in the experiment setup. In order to empirically prove the correctness of our implementation we compare the product classification task with the problem of spam categorization. Following that we describe our improved baseline method and the corresponding results and analysis in the \textit{Improved Baseline} section. The \textit{Discussion} section contains an overview of the problem domain and describes the evolution of our approach to the problem over time and the steps that lead us to devise and implement the proposed method. Finally we summarize the contributions of the paper in the \textit{Conclusion}.

\section{Background}
The problem of extracting product information from web and more classic corpora has been widely addressed in literature. Research in this area seems though be focused on documents that are known to represent a product or a brand like pages from web shops, news articles regarding items, fora and social networks which users discuss about selected topics. Our aim in is to extend the scope of the search in a general purpose domain like WikiPedia, which is a corpus composed of general topics and discussions, a fraction of which are actually products and brands. \\

In \textit{Deriving Marketing Intelligence from Online Discussion} \cite{1081919} the authors address the problem of extracting sentiment and opinions about products (PDAs in this case). The authors perform their analysis on a broad social network comprising weblogs, internet fora and usenet. An inspiring subtask discusses in this paper is topic detection, that appears to be similar to our brands/products discovering task.
An interesting approach the author propose is \textit{normalization} of extracted entities and the application of machine learning techniques to classify products. The authors suggest a classifier based on  Winnow, that according to the paper should outperform state-of-the-art methods such as SVM and KNN.
The key idea of this algorithm is to provide a linear separator between in-topic and off-topic documents. The authors propose a POS tagger for polarity discovery.

In the paper \textit{Comparative Experiments on Sentiment Classification for Online Product Reviews} \cite{1597389} the authors
 focus on tracking reviews to determine sentiments. Four classifiers (for sentiments) are described:
\begin{itemize}
\item PA
\item Winnow
\item Language Model based
\item High order N-grams are used as features in discerning sentiment.
\end{itemize}

\textit{Object-level Vertical Search} \cite{Nie2007Objectlevel} describes an object-level search paradigm in contrast to the usual page-level search paradigm. Particularly interesting for our work could be section 3. The paper deals with the problem of identifying products from a vast range of user generated content (with multiple templates). To our domain (WikiPedia) it is important to note that theoretically we have only one template, but in practice many differences may occur between wiki pages. We need to highlight common features; We don't have notions about price that would be a very useful indicator. Moreover, we often miss info about address, email and phone number. The authors propose an extraction method based on conditional random fields.

\section{Dataset: Creation and Evaluation}
Products are defined by the TREC guidelines as \textit{the most specific object that has a separate page under its manufacturer’s site} \footnote{http://ilps.science.uva.nl/trec-entity/guidelines/}. We aimed at extracting pages from wikipedia in a way that they could reasonably match this criteria; our method is generalized to recognize \textit{brand pages} as valid istances. Brand pages are defined either as pages describing a manufacturer (ie: Nike) or a line of products (ie: iPod).

To our knowledge at the moment of writing no known, freely accessible, dataset of product pages, extracted from WikiPedia, existed.
A crucial and time consuming task was to build such a set and setup a environment to test our method.
In this section we present the dataset used to compute the results presented in this paper
and the experiment setup.
The final approach we used for creating such a dataset is the result of an evolution over time
in the methodologies we considered to address product recognition.
In the \textit{Discussion} section we will present other methodologies we took into account 
and the reasons that lead us to the one actually employed.

\subsection{Creation}
We used the kaboodle \footnote{http://www.kaboodle.com/} product search engine to extract information from wikipedia. We implemented a web crawler to download links to pages reported as products and after having manually polished the list we have been able to obtain enough pages to attempt statistical analysis.
The output of this search engine was not perfect though and some manual interaction was needed to remove duplicate pages and false positives.

At this stage we focused only on English pages discarding documents in other languages.

Starting from a total set of approximately 4000 pages we were able to identity 679 English product pages by manually going through the collection and discarding
non English pages, duplicates and non product pages.

\subsection{Experiment Setup and Evaluation}
We evaluated the method by training against a set of 400+400 product and non product pages randomly chosen from the collection and testing
against a set composed of 195 + 195 product/non-product pages randomly chosen such that they do not appear in the training set. We will refer to this set as \textbf{set1}.

We ran our classifiers (both the baseline and the improved methods, which are described in details in the following sections) on 5 different typologies of experiment:

\begin{itemize}
\item Exp1: Training using the whole text of each page;
\item Exp2  Training using the whole text of each page plus terms extracted from the category list of that page;
\item Exp3: Training using the first 50 words of each page;
\item Exp4: Training using the first 50 words of each page plus terms extracted from the category list of that page;
\item Exp5: Training using only terms extracted from the category lists of pages;
\end{itemize}

Further, to prove the correctness of our implementation, we introduced two error correction experiments:
\begin{itemize}
	\item We trained and tested our method to distinguish spam emails from ham
	\item We tested product categorization against a set of known product pages manually selected from wikipedia and not present in the training collection. We made sure  that almost no overlap in the nature of products existed between the two collections. We will refer to this set as \textbf{set2}.
	
\end{itemize}

The metrics used for evaluating the methods are accuracy, precision and recall.
In the classification context those are defined according to the \textit{Accuracy Matrix} depicted in \textit{Table 1}

\begin{table}[!ht]
	\begin{center}
	
    \begin{tabular}{  c | c | c  }
      & \textit{product}  & \textit{non product}   \\
    \hline \hline
      \textit{product} & true positive (tp)  & false positive (fp)  \\
    \hline
      \textit{non product} & false negative (fn)  & true negative (tn)\\
    \end{tabular}
	\end{center}
     \caption{Accuracy matrix. The columns represent correct result/classification. The rows represent the obtained result/classification.}
\end{table}

The terms \textit{true positive}, \textit{true negative}, \textit{false positive} and \textit{false negative} are used to compare the given classification of an item (the class label assigned to that item) with the desired correct classification. The evaluation metrics are then defined as:

\begin{itemize}
	\item Accuracy: $\frac{tp + tn}{tp + tn + fp + fn}$
	\item Precision: $\frac{tp}{tp + fp}$
	\item Recall: $\frac{tp}{tp + fn}$
\end{itemize}

\section{Classification}
In order to perform product (brand) recognition from wikipedia pages we used a Naive Bayes Classifier(NBC) \cite{citeulike:873540} trained to classify a page as product or brand given a feature set extracted from given product pages. At this stage we did not focus on multilingual tracking, aiming only at English candidate pages. In this section an introduction to the methods used for the baseline will be presented as well as the results obtained.

\subsection{Baseline}
Pages are represented as \textit{unigram language models} and a Naive Bayes Classifier(NBC) with a TF-IDF metric is applied to achieve the goal.
In this section we first introduce the theoretical fundaments of Language Models, Naive Bayes Classifiers(NBCs) and TF-IDF and we then present and discuss
results obtained on a baseline implementation and on its improvement. Error correction has been carried on to prove the correctness of our implementation.
The dataset and experiment setup used to obtain these results are the ones described in the previous section.

\subsubsection{Method}
The method consist of four components:
\begin{itemize}
	\item A representation of documents as language models
	\item A classifier
	\item A metric to weight the meaning of words and assist classification
	\item A learning and classification procedure
\end{itemize}

\paragraph{Language Models\\}
A statistical language model \cite{wikipedia:lm} assigns a probability $P(w_1,...,w_m)$ to a sequence of $m$ words  by means of a probability distribution.
In a unigram language model this probability is approximated as:
$P(w_1,...,w_m) = \prod_{i=1}^m P(w_i)$.

We used \textit{unigram language models} to represent product and non-product pages. Two separate models have been built from the training set so to represent the two different kinds of documents.

In the presented model features of the classes, in terms of Naive Bayes Classification (next section), are words occurring in product and non product documents. The probability of each word $w_i$ given that it belongs to a class $C_k$ is approximated with \textit{relative frequencies} from the training set.
\begin{equation}
	P(w_i | C_k) = \frac{n_{C_k}(w_i)}{\sum_i n_{C_k}(w_i)}
\end{equation}

where $n_{C_k}(w_i)$ is the number of occurrencies of word $w_i$ in class $C_k$ of the training set and this count is normalised over the total number of word occurring in the set. These probabilities are maximum likelihood estimates of the probabilities.

\paragraph{Naive Bayes Classifier\\}

Naive Bayes is a classification that employs Bayes formula with strong independence assumptions. Bayes formula states that $P(A|B) = \frac{P(B|A) \cdot P(A)}{P(B)}$, where
\begin{itemize}
	\item $P(A)$ is the \textit{prior} probability or \textit{marginal} probability of A, in a sense that it does not take into account any information about B.
	\item $P(A|B)$ is the \textit{conditional} probability of A given B.
	\item $P(B|A)$ is the \textit{conditional} probability of B given A.
	\item $P(B)$ is the \textit{prior} or marginal probability of B
	
\end{itemize}

\textit{A naive Bayes classifier assumes that the presence (or absence) of a particular feature of a class is unrelated to the presence (or absence) of any other feature. 
For example, a fruit may be considered to be an apple if it is red, round, and about 4" in diameter. Even though these features depend on the existence of the other features, a naive Bayes classifier considers all of these properties to independently contribute to the probability that this fruit is an apple.} \cite{wikipedia:nbc}.

For our approach the model for an NBC is a conditional probabilistic model over a class variable $C_k$ and a set $F_1...F_n$ of features:
\begin{equation}
	P(C_k|F_1,...F_n) = \frac{P(F_1,...F_n|C_k)\cdot P(C_k)}{P(F_1,...,F_n)}
\end{equation}

The denominator part of (1) can be discarded because it will remain constant for all given classes and serves as a scale factor. What we are interested is maximizing the likelihood of the nominator. The problem we aim to solve, classify wikipedia pages, is a two class (boolan) task. A first class is given by \textit{product} (or \textit{brand}) pages
whereas the second consists in \textit{non-product} (or \textit{non-brand}) pages. Features charachterizing a class are given by words occuring respectively
in product and non-product pages.

Using the unigram model, the probability of a set of features given a class can be estimated as:
\begin{equation}
	P(F_1,..,,F_n | C_k) = \prod_{i=1}^n P(F_i | C_k)
\end{equation}

All model parameters (class priors and feature probability distributions) can be approximated with relative frequencies from the training set. These are \textit{maximum likelihood estimates} of the probabilities.

From the probability we can build a classifier by defining a function like:
\begin{equation}
	classify(f_1,...,f_n) = argmax_{C_k} P(C_k) \prod_{i=1}^n P(F_i = f_i | C_k).
\end{equation}

which can be described as: \textit{a document represented by a given set of features f\_i...f\_n is classified as belonging to a class $C_k$ (product or non product) such that $C_k$ is the most-probable class}. This decision rule is known as the \textit{maximum a posteriori} or MAP choice. NBCs are called \textit{naive} because of their conditionally independence assumption between features, given the class of a document. As we mentioned in the previous section, we used two separate models to represent the two different kinds of documents. Now that we have seen how the NBC works, we need to find features for each document.

\paragraph{Ranking of Words\\}

In order to rank words in the two classes, given their estimated probability, we borrowed a ranking criteria often used in vector space representation \textit{term frequency-inverse document frequency} (TF-IDF) \cite{wikipedia:tfidf}. For classification we select the top $n$ words given their rank and use them as features. The \textit{inverse document frequency factor} is incorporated which diminishes the weight of terms that occur very frequently in all the pages in the collection regardless of their class and increases the weight of terms that occur rarely. To explain this choice we have to remind that the goal of our classifier is to categorize a given (unseen in the training collection) page either as product (brand) or non-product (non-brand). Simply counting the frequency of each word in product and not product pages is not a good heuristic; even after having performed stopwords removal, stemming and text normalization we encountered difficulties in properly being able to characterize pages. This weight enforced by TF-IDF on terms is a statistical measure used to evaluate how important a word is to a document in a collection or corpus and is defined as:

\begin{equation}
	tf_{ij} = \frac{n_{i,j}}{\sum_k n_{k,j}}
\end{equation}

where $n_{i,j}$ is the number of occurrences of the considered term $t_i$ in document $d_j$, and the denominator is the sum of number of occurrences of all terms in document $d_j$.

\begin{equation}
	idf_{i} = \log \frac{|D|}{|\{d : {t_i \in D}\}|}
\end{equation}

where $|D|$ is the total number of the documents and the denominator represents the number of documents in which term $t_i$ appears.

In vector spaced tf-idf is then defined as:
\begin{equation}
	tf-idf_{i,j} = tf_{i,j} \times idf_i
\end{equation}

Given that we know the number of terms of our data set we borrowed the tf-idf underlying idea and introduced an \textit{inverse-frequency} to rank the terms in our language model. In particular we wanted to adjust weights taking to highlight words that:
\begin{itemize}
	\item appear frequently in a single product page
	\item appear rarely but in multiple product pages
\end{itemize}

\paragraph{Classification\\}
The learning approach can be summarized by the pseudocode depicted in \textit{Figure 1}.

\begin{figure}[h]
	\begin{verbatim}
      prod = list of procut_pages
      non-prod = list non-product_pages
      
      train(prod, non-prod):
         # initialize language models
         lm_prod = Nil 
         lm_non-prod = Nil
         # create language models for the collection
         for p in prod:
            update(lm_prod, p)     # embed TF-IDF information
         for np in non-prod:
            update(lm_non-prod, np) # embed TF-IDF information
         
         return lm_prod, lm_non-prod
      end
	\end{verbatim}
	\caption{Model training phase}
\end{figure}

In which the update function, updates the corresponding language model using the new product (or non-product) page (The TF-IDF ranking is included in the update fuction, so at the end we have the features list with the highest ranks.). An unseen page is classified by first building a language model for it and performing a Maximum a Posteriori choice following the
definition of (2) and comparing against the collection language models(\textit{Figure 2}). 

\begin{figure}[h]
	\begin{verbatim}
      classify(page):
        lm = build_language_model(page)
        # Do a Map choice as in formula(2)
        # using the language models that were computed in training
        c = classify(page)          
        return c
      end
	\end{verbatim}
   \caption{Classification phase}
\end{figure}

\subsubsection{Results}

We evaluated the method by training against a set of 400+400 product and non product pages randomly chosen from the collection and testing
against a set composed of 195 + 195 product/non-product pages randomly chosen such that they do not appear in the collection (set1).

We ran the classifier (both baseline and improved) on 5 different typologies of experiment as described in \textit{Section 3.2}.

\begin{table}[ht]
	\begin{center}
	
    \begin{tabular}{  c | c }
     NBC & \textbf{Accuracy} \\
    \hline \hline
      \textit{Exp1} & 0.477 \\
    \hline
      \textit{Exp2} & 0.477\\
    \hline
      \textit{Exp3} & 0.5  \\
    \hline
      \textit{Exp4} & 0.5  \\
    \hline
      \textit{Exp5} & 0.5  \\
    \end{tabular}
	\end{center}
     \caption{Accuracy brand/product classification, $P(product) = P(non\_product) = \frac{1}{2}$ on set1}
\end{table}

\begin{table}[ht]
	\begin{center}
	
    \begin{tabular}{  c | c}
     NBC & \textbf{Accuracy}  \\
    \hline \hline
      \textit{Exp1} & 0.477 \\
    \hline
      \textit{Exp2} & 0.477  \\
    \hline
      \textit{Exp3} & 0.5  \\
    \hline
      \textit{Exp4} & 0.5  \\
    \hline
      \textit{Exp5} & 0.5  \\
    \end{tabular}
	\end{center}
     \caption{Accuracy brand/product classification, $P(product) = \frac{1}{3}, P(non\_product) = \frac{2}{3}$ on set1}
\end{table}

\subsubsection{Analysis}

The results show some problems of applying the baseline method to the given data set(\textit{Table 2}). First we noticed that when the full 
context of a page is used, the classifier is biased to recognizing all new instances as products. In order to mitigate this 
problem we performed a second run of experiment changing the prior probabilities of the documents. In the first run we assumed the probability $P(product) = P(non\_product) = \frac{1}{2}$. This is not realistic because in the real case scenario we expect more non product pages than product ones. In the second run we adjusted the probabilities to $P(product) = \frac{1}{3}$ and 
$P(non\_product) = \frac{2}{3}$. We did so to resemble the ratio of product and non product pages of the corpus we sample pages to use for training and testing. Again we obtain the same results in both cases(\textit{Table 3}). \\
Another experiment we performed was to use \textit{wikipedia's category} words to better characterize a page. We did so in two ways:
we first used text extracted from the page to which we added categories and we then used categories only to perform classification. The results we obtained are not particularly meaningful and close to random choice.
Looking at domains where NBC proved to be a strong solution we think that part of the problem resides in kind of data we are trying to analyze. \textit{Spam classification} is a domain were NBC is considered a strong classifier \cite{sahami98bayesian}; in that case it is though possible
to characterize emails given unique features (words) that are likely appear with a high frequency in spam emails whereas they are not so common in ham emails. For this reason a language model built using spam content for training will be different (in terms of words and frequencies) from one built using ham. As a consequence new instances are more likely to fit one of the two models better.
In our case there is a great overlap of words among product and non products pages, this leads to having language models with very close word probabilities. The result is that once classification is attempted a new instance is likely to fit equally good one of the two models, thus resulting a classification similar to a random choice. \\
NBCs and language models were used in literature we analyzed as a starting point for our project. The domains were this methods have been attempted where much narrow than the wikipedia corpus. For instance, if we want to train a classifier to recognize PDA
products we can focus on words appearing only in pages describing PDAs (ie: \textit{battery life}, \textit{resolution}, \textit{personal assistant manager}) and brands that are known to produce PDAs. Current research also focuses on domain specific corpora; sentiment analysis and text classification of brands or products is usually performed on text extracted from webshops, magazines, and manufacturer sites that deal with a given kind of products; in this case it is possible to extract features like price, manufacturing date and etc in a more structured and consistent way (for example by analyzing the html code to extract description boxes) whereas in wikipedia this features are both often missing and the structure of pages is not uniform among each other.

We mentioned the goodness of NBCs in spam/ham classification. As a way to compare and better understand our results we run our classifier against a collection of spam and ham emails with the aim of recognizing new instances of spam emails.

\begin{table}[ht]
\begin{center}
	\begin{tabular}{  c | c | c  | c  | c}
      NBC & \textbf{Accuracy}  & \textbf{Precision}  & \textbf{Recall}  \\
     \hline \hline
       \textit{Spam} & 0.886  & 0.956  & 0.815  \\
	\end{tabular}
\end{center}
	\caption{Spam classification}
\end{table}

\textit{Table 4} depicts the results for spam classification. Training has been performed on a set of 182 spam and 226 ham emails. For testing 45 spam and 145 ham emails have been used. Priors have been set so that $P(spam) = \frac{2}{3}$ and $P(non\_spam) = \frac{1}{3}$ and all words present in the language models built during the training phase have been taken into consideration as possible features. These results are similar to the ones we previously reported in a previous work on spam/ham classification tasks and show that our implementation is correct. Reasons for better performance can be found in the characteristic of the spam/ham text classification domain described above.

\subsection{Improved Baseline}
As an improvement over the baseline method we aimed at characterizing pages by extracting words highly frequent in products (not the words that occur many times in a couple of product pages) and not in non-products and vice versa. In the baseline we used the term frequency as the nominator for the words (features) ranking method. After analyzing the results in the improved baseline we used the document frequency instead of term frequency. This is because we found that there are many words occuring many times in just a couple of product(or non-product) pages, so they are not good features for product (or non-product) pages. Therefore using document frequency helps to find the features that are most informative for each class (products or non-products). For instance the word "released" may occur in many product pages, but in each one just once. So by using document frequency we try to find such words that are generally usefull as a feature for products or non-products.

On top of that we performed manual analysis to better select features. Our goal was to determine if using a less number of very meaningful words would have had an impact on the correctness of the classifier. \\
As a further improvement we introduced \textit{Laplace smoothing} on the relative frequency estimate of words computed during the training phase to reserve probability mass for terms occurring with null probability in the test set.
For a given word $w_i$, the smoothed $P(w_i | C_k)$ probability has been estimated as:
\begin{equation}
	P(w_i | C_k) = \frac{n_{C_k}(w_i) + 1}{\sum_i n_{C_k}(w_i) + |V|}
\end{equation}
where $|V|$ is the features set size, which is equal to the vocabulary size in case we use all the words as features.

We ran the improved classifier under the same experiment setup as baseline (same experiments and same training/test sets).

This new approach leads to better and more promising results as can be seen in Table 5 and 6.

	\begin{table}[!ht]
		\begin{center}
		
        \begin{tabular}{  c | c | c  | c  | c}
         NBC & \textbf{Accuracy}  & \textbf{Precision}  & \textbf{Recall}  \\
        \hline \hline
          \textit{Exp1} & 0.704& 0.894 & 0.750 \\
        \hline
          \textit{\textbf{Exp2}} & \textbf{0.717} & \textbf{0.902} & \textbf{0.766}\\
        \hline
          \textit{Exp3} & 0.443 & 0.710 & 0.396 \\
        \hline
          \textit{Exp4} & 0.689 & 0.882& 0.739 \\
        \hline
          \textit{Exp5} & 0.685 & 0.817 & 0.817 \\
        \end{tabular}
		\end{center}
         \caption{Accuracy, precision and recall of brand/product classification on set1}
	\end{table}
	
	\begin{table}[!ht]
		\begin{center}

	    \begin{tabular}{  c | c | c  | c  | c}
	     NBC & \textbf{Accuracy}  & \textbf{Precision}  & \textbf{Recall}  \\
	    \hline \hline
		  \textit {Exp2 100ft} & 0.566 & 0.792 & 0.614 \\
		 \hline
	      \textit{Exp2 200ft} & 0.612  & 0.792  & 0.713  \\	
	   	 \hline
	      \textit{\textbf{Exp2 500ft}} & \textbf{0.639}  & \textbf{0.801} & \textbf{0.755} \\	
		 \hline
		  \textit {Exp4 100ft} & 0.545 & 0.878 & 0.489 \\
		 \hline
	      \textit{Exp4 200ft} & 0.557  & 0.876  & 0.515 \\
		 \hline
	      \textit{\textbf{Exp4 500ft}} & \textbf{0.596} & \textbf{0.870}  & \textbf{0.594}  \\
		 \hline
		  \textit {Exp5 100ft} & 0.609 & 0.917 & 0.578 \\
		 \hline
		  \textit {Exp5 200ft} & 0.616 & 0.865 & 0.635 \\
		 \hline
		  \textit{\textbf{Exp5 500ft}} & \textbf{0.637} & \textbf{0.856} & \textbf{0.682} \\
	    \end{tabular}
		\end{center}
	     \caption{Accuracy, precision and recall of brand/product classification on set1 (using three different number of features for each type of experiment)}
	\end{table}
	
	\newpage
	
\subsubsection{Analysis}
Table 5 shows the result of the improved baseline on the proposed experiments. The table shows improvements in the evaluation metrics. These results suggest the importance of finding a good and balanced correlation between words describing product and non product pages. 

Table 6 depicts the results of running experiments and using only a subset of words as classification features. For classification we selected the top $n$ words given their tf-idf rank and used them as features. Results are worse than the ones obtained in the first experiment and we can see that performance tends to increase by increasing the number of features.
It is important to note that precision and recall seem to be affected by the number of features employed. Precision decreases when a higher number of features is used, while recall increases. This suggests that number of features is a parameter that should be tuned given a domain specific task in a way to favor one of the metrics (for instance \textit{Exp5} in Table 5 has the highest recall which means using only categories leads to a better recall than other experiments).

As a further error correction methodology we tested our method on a set of 151 pages (set2) not present in the set extracted from kaboodle for the experiments described before (set1). These pages have been extracted from the wikipedia \textit{List of Ebooks} \footnote{http://en.wikipedia.org/wiki/List\_of\_e\-book\_readers} for products and \textit{LVMH} \footnote{http://en.wikipedia.org/wiki/LVMH} for brand pages.
Table 7 shows the results for our improved baseline method obtained on the error correction set.
\begin{table}[ht]
\begin{center}
	\begin{tabular}{  c | c }
      NBC & \textbf{Accuracy}  \\
     \hline \hline
       \textit{Exp2} & 0.349 \\
	 \hline
		\textit{\textbf{Exp5}} & \textbf{0.655} \\
	\end{tabular}
\end{center}
	\caption{Error correction against a set of known brands/products (set2)}
\end{table}

Our training collection, as described, has been extracted using the kaboodle search engine. Products retrieved belong mostly
to multi media, videogames, and literature products. The products present in set 2 are very different in nature. For instance
a lot of references to wines and watches are found while we almost have no notion of them in the training set.
We did so to test our method on a more general setup. Given that we know that what we are going to classify only product pages. Performance has been estimated in terms of accuracy. Given the nature of the dataset, which is comprised of product pages only, this value is equivalent to recall. Exp2 and Exp5 have been chosen since they are the two that present the best results in terms of recall. In case of Exp2 we assist to a drop of performance, while with Exp5 proves to be still better than random choice. Once again this seems to show the importance of using category terms in classification when we want to tune an application towards recall.

\subsection{Discussion}
In order to achieve our goal we attempted various strategies and we analyzed the problem from different viewpoints.We began by looking for a definition of product and a way to model products as entities. Products are defined by the TREC Entity track guidelines as \textit{the most specific object that has a separate page under its manufacturer’s site} \footnote{http://ilps.science.uva.nl/trec-entity/guidelines/}. We aimed at extracting pages from wikipedia in a way that they could reasonably match this criteria.

In order to perform a case study we needed a list of products extracted from the wikipedia corpus. The first approach consisted of focusing on \textit{brands} and trying to exploit wikipedia categories and \textit{list of} pages to gain useful information. We attempted text mining both on the dump provided by the wikimedia foundation \footnote{http://en.wikipedia.org/wiki/Wikipedia\_database} and a list of categories provided by the INEX benchmark \footnote{http://www.inex.otago.ac.nz}.

With this approach we had to face two main problems. The list of categories available for the TREC task contains a lot of unmet references in the current version of wikipedia that required a human effort to be solved. At the same time, the semi-structured nature of wikimedia made it difficult to write a bias free crawler able to extract categories and lists from the SQL dump. At this stage we identified two alternatives to extract product and brand pages: rule based and statistical learning.

Given time constraints and lack of training data we decided to attempt a rule based approach to extract information from wikipedia; the problem at this point was defining rules general enough to capture any possible kind of products. Rule based approach requires a high level of human interaction to hard code patterns and has the drawback of not being very scalable. While searching for pages to analyze and extract recurring patterns we found ourselves often biased by our own interests; Plus, given that wikipedia is a container of user contributed contents, patterns may vary from page to page and from product to product. The lack of generality and the time required to effectively extract patterns and hardcode rules forced us to look for other solutions.

An alternative to rule based learning is statistical modelling; in order to perform this type of learning training data is needed to extract frequency count of words and other probabilistic information. This approach, despite being promising (see references), lead us to a circular dependency. On one side we wanted to use statistical information to discover new products, on the other hand we needed a set of product pages extracted from wikipedia to perform training.

To solve this problem of creating a training set we looked for two possible solutions.

First we tried to exploit the ontologies provided from the DBpedia \footnote{http://dbpedia.org/} project to obtain a better refined view of wikipedia categories. The ontology collection though is not focused on product/brands and this path soon lead us to the very same problems faced at the beginning (a hugh human effort to clean up categories).

The second solution we adopted, the one chosen for our baseline experiment, is to use the kaboodle \footnote{http://www.kaboodle.com/} product search engine to extract information from wikipedia. We implemented a web crawler to download links to pages reported as products and after having manually polished the list we have been able to obtain enough pages to perform the probabilistic analysis described in this paper.

\section{Conclusion}
In this paper we described the problem of extracting product and brand pages from the wikipedia corpus. Several approaches have been attempted that lead us to model the problem as a classification task. An important contribution of our work consists of the creation of a dataset of selected product/brand pages extracted from wikipedia and a related experiment setup. We described a baseline approach based on Naive Bayes classification. After having highlighted some problems that arose with this method we proposed and improvement that actually lead to better results in terms of accuracy, precision and recall. Finally we performed error correction on our method by applying it to the spam/ham classification domain and testing on a separate set of known brand/product pages.

\newpage
\section{Appendix}
\subsection{Most informative words for brands and non brands}
\subsubsection{Brand Features}

\begin{table}[h]
\begin{center}
	\begin{tabular}{  c | c | c | c }
      Word & Term frequency in brands & Document frequency in brands & Document frequency\\
    \hline \hline
       \textit{bwv} & 952 & 2 & 2 \\
	 \hline
       \textit{film} & 5432 &  216 & 333 \\
	 \hline
       \textit{iphon} & 1040 & 7 &  9 \\
	 \hline
       \textit{game} & 3257 & 146 &  240 \\
	 \hline
       \textit{appl} & 1096 &  46 &  75 \\
	 \hline
       \textit{season} & 1747 &  101 &  190 \\
	 \hline
       \textit{episod} & 1605 &  114 &  171 \\
	 \hline
       \textit{wii} & 684 & 26 &  28 \\
	 \hline
       \textit{seri} & 2528 & 203 & 364 \\
	 \hline
       \textit{acacia} & 372 & 2 & 5 \\
	 \hline
       \textit{cola} & 425 & 5 & 10 \\
	 \hline
       \textit{nintendo} & 609 & 38 & 42 \\
	 \hline
       \textit{movi} & 1592 & 191 & 259 \\
	 \hline
       \textit{guitar} & 636 & 32 & 49 \\
	 \hline
       \textit{album} & 1067 & 99 & 154 \\
	 \hline
       \textit{csi} & 407 & 7 & 11 \\
	 \hline
       \textit{playstat} & 567 & 32 & 37 \\
	 \hline
       \textit{ign} & 695 & 62 & 68 \\
	 \hline
       \textit{award} & 1483 & 158 & 251 \\
	 \hline
       \textit{ikea} & 283 & 1 & 2 \\
	 \hline
       \textit{tardi} & 301 & 2 & 3 \\
	 \hline
       \textit{releas} & 2812 & 284 & 438 \\
	 \hline
       \textit{2007} & 4116 & 275 & 529 \\
	 \hline
       \textit{2008} & 5088 & 298 & 572 \\
	 \hline
       \textit{player} & 1188 & 119 & 196 \\
	\end{tabular}
\end{center}
\end{table}
\newpage
\subsubsection{Non-Brand Features}

\begin{table}[h]
\begin{center}
	\begin{tabular}{  c | c | c | c }
      Word & Term frequency in non-brands & Document frequency in non-brands & Document frequency\\
    \hline \hline
       \textit{irv} & 761 & 12 & 23 \\
	 \hline
       \textit{soviet} & 899 & 35 & 44 \\
	 \hline
       \textit{mandela} & 442 &  4 &  5 \\
	 \hline
       \textit{citi} & 2256 & 173 & 303 \\
	 \hline
       \textit{glutam} & 349 &  1 & 2 \\
	 \hline
       \textit{hdmi} & 446 & 2 & 8 \\
	 \hline
       \textit{laker} & 307 & 1 & 1 \\
	 \hline
       \textit{govern} & 1472 & 142 & 203 \\
	 \hline
       \textit{calla} & 332 & 1 & 2 \\
	 \hline
       \textit{church} & 941 & 81 & 103 \\
	 \hline
       \textit{olymp} & 646 & 39 & 48 \\
	 \hline
       \textit{nation} & 1922 & 201 & 323 \\
	 \hline
       \textit{vietnam} & 623 & 38 & 52 \\
	 \hline
       \textit{ottoman} & 405 & 12 & 13 \\
	 \hline
       \textit{tiger} & 527 & 22 & 37 \\
	 \hline
       \textit{moscow} & 491 & 24 & 31 \\
	 \hline
       \textit{ogg} & 531 & 23 & 42 \\
	 \hline
       \textit{utc} & 474 & 27 & 30 \\
	 \hline
       \textit{msg} & 515 & 22 & 40 \\
	 \hline
       \textit{popul} & 856 &  104 & 134 \\
	 \hline
       \textit{puerto} & 424 & 17 & 22 \\
	 \hline
       \textit{iran} & 515 & 37 & 45 \\
	 \hline
       \textit{hitler} & 403 & 17 & 21 \\
	 \hline
       \textit{isbn} & 1615 & 186 & 321 \\
	 \hline
       \textit{bbc} & 875 & 75 & 152 \\
	\end{tabular}
\end{center}
\end{table}
\newpage

\subsection{Most informative words for brands and non brands when category terms are used}
\subsubsection{Brand Features}

\begin{table}[h]
\begin{center}
	\begin{tabular}{  c | c | c | c }
      Word & Term frequency in brands & Document frequency in brands & Document frequency\\
    \hline \hline
       \textit{films} & 630 & 85 & 91 \\
	 \hline
       \textit{games} & 236 & 34 &  35 \\
	 \hline
       \textit{series} & 235 & 55 & 61 \\
	 \hline
       \textit{Films} & 235 & 71 & 75 \\
	 \hline
       \textit{television} & 208 & 52 & 69 \\
	 \hline
       \textit{statements} & 379 & 138 & 300 \\
	 \hline
       \textit{with} & 475 & 189 & 373 \\
	 \hline
       \textit{unsourced} & 347 & 135 & 293 \\
	 \hline
       \textit{Television} & 111 & 33 & 38 \\
	 \hline
       \textit{by} & 152 & 105 & 116 \\
	 \hline
       \textit{in} & 246 & 139 & 249 \\
	 \hline
       \textit{American} & 162 & 98 & 141 \\
	 \hline
       \textit{from} & 493 & 220 & 465 \\
	 \hline
       \textit{2009} & 273 & 149 & 304 \\
	 \hline
       \textit{Articles} & 461 & 221 & 454 \\
	 \hline
       \textit{needing} & 173 & 104 & 209 \\
	 \hline
       \textit{set} & 90 & 60 & 62 \\
	 \hline
       \textit{novels} & 73 & 37 & 38 \\
	 \hline
       \textit{the} & 148 & 81 & 180 \\
	 \hline
       \textit{video} & 73 & 37 & 41 \\
	 \hline
       \textit{2008} & 165 & 119 & 230 \\
	 \hline
       \textit{software} & 58 & 17 & 24 \\
	 \hline
       \textit{articles} & 327 & 217 & 429 \\
	 \hline
       \textit{albums} & 52 & 15 & 18 \\
	\end{tabular}
\end{center}
\end{table}
\newpage
\subsubsection{Non-Brand Features}

\begin{table}[h]
\begin{center}
	\begin{tabular}{  c | c | c | c }
      Word & Term frequency in non-brands & Document frequency in non-brands & Document frequency\\
    \hline \hline
       \textit{of} & 449 & 144 & 207 \\
	 \hline
       \textit{statements} & 457 & 162 & 300 \\
	 \hline
       \textit{unsourced} & 434 & 158 & 293 \\
	 \hline
       \textit{with} & 541 & 184 & 373 \\
	 \hline
       \textit{from} & 615 & 245 & 465 \\
	 \hline
       \textit{Articles} & 585 & 233 & 454 \\
	 \hline
       \textit{the} & 217 & 99 & 180 \\
	 \hline
       \textit{in} & 236 & 110 & 249 \\
	 \hline
       \textit{2009} & 285 & 155 & 304 \\
	 \hline
       \textit{United} & 120 & 57 & 92 \\
	 \hline
       \textit{States} & 112 & 52 & 81 \\
	 \hline
       \textit{needing} & 190 & 105 & 209 \\
	 \hline
       \textit{Birds} & 57 & 10 & 10 \\
	 \hline
       \textit{English} & 67 & 24 & 26 \\
	 \hline
       \textit{articles} & 346 & 212 & 429 \\
	 \hline
       \textit{American} & 123 & 43 & 141 \\
	 \hline
       \textit{containing} & 95 & 57 & 89 \\
	 \hline
       \textit{pages} & 119 & 86 & 143 \\
	 \hline
       \textit{All} & 310 & 212 & 412 \\
	 \hline
       \textit{and} & 110 & 85 & 126 \\
	 \hline
       \textit{2008} & 160 & 111 & 230 \\
	 \hline
       \textit{involving} & 40 & 6 & 6 \\
	 \hline
       \textit{Wikipedia} & 116 & 84 & 154 \\
	 \hline
       \textit{text} & 76 & 47 & 68 \\
	 \hline
       \textit{language} & 74 & 45 & 65 \\
	\end{tabular}
\end{center}
\end{table}

\bibliographystyle{plain}
\bibliography{lib}

\end{document}